\documentclass[reprint,
twocolumn,
amsmath,
amssymb,
aps,
physrev]{revtex4-2}
\usepackage{graphicx}
\usepackage{dcolumn}
\usepackage{bm}
\usepackage{lineno} 
\usepackage{comment} 
\usepackage{multirow}
\usepackage{hyperref} 
\hypersetup{ 
	colorlinks=true,
	linkcolor=blue,
	filecolor=magenta,      
	urlcolor=blue,
	citecolor=red
}
\usepackage{labelschanged}
\usepackage{xcolor} 
\usepackage{amsfonts, amssymb, bbm}
\usepackage[bb=libus]{mathalpha}

\begin{document}
	\title{\textbf{Helicity signatures in Schwinger pair production under rotating electric fields}}
	
	\author{Abhinav Jangir}
	\email{2022rpy9087@mnit.ac.in}
	\affiliation{Department of Physics, MNIT Jaipur, Jaipur, Rajasthan, India}
	\date{\today}
	
	\begin{abstract}
	We investigate helicity effects in electron--positron pair production in circularly polarized electric fields with frequency chirping using the quantum-kinetic formalism for fermions. In particular, we calculate helicity-resolved momentum distributions and characterize the helicity imbalance through the helicity-asymmetry parameter for various electric-field configurations, including temporally asymmetric fields and dynamically assisted two-color fields with different frequency-chirping configurations. We find that right- and left-helicity electrons preferentially populate opposite momentum half-spaces along the out-of-plane momentum direction, leading to a pronounced helicity asymmetry in the produced electron distributions. For temporally asymmetric fields, frequency chirping substantially modifies the momentum distributions and can strongly enhance the helicity asymmetry, with the response depending sensitively on the pulse asymmetry parameter. For symmetric and elongated pulses, the helicity asymmetry approaches $100\%$ at sufficiently large chirp strengths, whereas the compressed-pulse case exhibits a qualitatively different response with suppressed asymmetry at large chirp. For dynamically assisted fields, chirping the weak high-frequency component produces the strongest helicity response, reaching a maximum asymmetry of nearly $87\%$ in the combined field, while chirping the strong field alone results in only a moderate asymmetry of about $12\%$--$14\%$. Simultaneous chirping produces a large helicity imbalance together with substantial restructuring of the momentum spectrum, although its helicity response is weaker than that obtained by chirping the weak field alone. These results demonstrate that helicity-resolved momentum distributions provide additional information on the spin-dependent dynamics of Schwinger pair production beyond that contained in helicity-summed spectra.
	\end{abstract}
	
	\maketitle
	\clearpage
	\section{Introduction}\label{sec:intro}
	The study of matter in extreme electromagnetic fields is a frontier of modern physics. While perturbative methods suffice for weak fields, high-intensity lasers and astrophysical fields approaching the Schwinger critical strength, \(B_{\mathrm{cr}}\simeq4.4\times10^{9}\,\mathrm{T}\) \cite{ge2020insight}, motivate nonperturbative approaches to strong-field quantum electrodynamics (SFQED). A key prediction is the creation of electron–positron (\(e^+e^-\)) pairs from the quantum vacuum, known as the Schwinger–Sauter effect, originating from the works of Sauter and Schwinger \cite{sauter1931behavior, schwinger1951gauge}. The characteristic electric-field scale is $E_{\rm cr}=m^2c^3/e\hbar\approx1.3\times10^{18}\,\mathrm{V/m}$ (where $m$ is the electron mass and $-e$ is the electron charge), above which pair production becomes appreciable. However, exponential suppression and the limited laboratory field strengths have so far prevented its direct observation. The theoretical framework was subsequently extended through the Euler–Heisenberg effective action \cite{heisenberg1936folgerungen} and the Green-function formulation of Nikishov \cite{nikishov1970pair}.
	
	Recent advances in ultra-intense laser technology \cite{pike2014photon, marklund2009quantum} have brought the experimental investigation of SFQED considerably closer to reality. Modern laser systems have achieved intensities of order $10^{22}\,\mathrm{W/cm^2}$ \cite{yanovsky2008ultra}, while next-generation facilities aim at $10^{25}-10^{26}\,\mathrm{W/cm^2}$ \cite{ringwald2001pair}, corresponding to field strengths approaching $E\approx0.1E_{\rm cr}$. The continued development of ultrashort, high-intensity laser facilities, such as the extreme light infrastructure \cite{ELI} and x-ray free-electron laser \cite{XFEL}, may provide access to the experimental observation of QED vacuum decay and offer a direct test of nonperturbative SFQED. Recent studies have explored a wide range of high-energy phenomena in intense laser fields, spanning relativistic quantum dynamics, SFQED, and nuclear and particle physics; see Refs. \cite{piazza2012extremely, fedotov2023advances, gonoskov2022charged} for comprehensive discussions. 
	
	Advances in computational methods have enabled the systematic exploration of nonperturbative vacuum pair production in strong electromagnetic backgrounds using a variety of theoretical frameworks. Among the widely employed approaches are semiclassical methods based on worldline instantons \cite{gies2005pair, dunne2006worldline, schneider2016dynamically, dumlu2011complex}, effective-action and effective-Lagrangian formulations \cite{dunne2005fields}, and the Furry-picture treatment of quantum fields in external backgrounds \cite{aleksandrov2020pair, aleksandrov2017momentum}. The dynamics of pair creation can also be studied within quantum kinetic theory, most notably through the quantum Vlasov equation \cite{kluger1998quantum, huet2014vlasov, alkofer2001pair, nuriman2012enhanced, abdukerim2013effects} and its equivalent Riccati-type formulation \cite{dumlu2010schwinger, marinov1977electron}. Other complementary approaches include computational quantum field-theoretical methods \cite{braun1999numerical, li2021study, krekora2004klein, tang2013electron}, S-matrix formulations \cite{muller2003differential, muller2003nonlinear, deneke2008bound}, and the Dirac-Heisenberg-Wigner (DHW) formalism \cite{blinne2014rotating, li2015effects}. Collectively, these approaches provide complementary perspectives on the nonperturbative dynamics of vacuum decay and particle production in SFQED.
	
	Over the past decades, extensive theoretical research has investigated vacuum pair production in temporally varying, spatially homogeneous electric fields, with particular emphasis on the temporal profile and carrier frequency \cite{hebenstreit2009momentum, dumlu2009equivalence, tanji2009dynamical, taya2021exact}. These studies were subsequently extended to spatially inhomogeneous fields, revealing the role of spatial structure in pair production \cite{hebenstreit2011particle, kohlfurst2018pondermotive, li2021study, li2021enhanced, ababekri2019effects, ababekri2020chirp, bake2024enhanced, bake2025vacuum, aleksandrov2016elelectron, aleksandrov2017dipole, aleksandrov2017momentum, aleksandrov2020pair}. Further investigations have explored complex temporal structures, including multipulse, sub-cycle and multicycle fields, as well as carrier-envelope phase and interference effects \cite{jangir2026carrier, abdukerim2013effects, aleksandrov2017shape, akkermans2012ramsey, dumlu2010schwinger, dumlu2011complex, dumlu2011interference, hebenstreit2009momentum}. 
	
	Since the field strengths required for an appreciable Schwinger pair-production rate remain beyond current experimental capabilities, considerable attention has been devoted to mechanisms that can enhance vacuum pair production in subcritical fields \cite{titov2012enhanced, schutzhold2008dynamically, dunne2009catalysis, piazza2009barrier, bulanov2010multiple}. Among these, the dynamically assisted Schwinger effect (DASE) provides a particularly effective mechanism by combining a strong, low-frequency field with a weak, high-frequency component, thereby substantially increasing the production rate relative to the corresponding single-field configuration \cite{schutzhold2008dynamically}. The efficiency of this enhancement is sensitive to the temporal profile of the assisting pulse \cite{linder2015}. DASE has therefore been extensively investigated for a wide range of field characteristics and parameters, considering both temporally and spatially inhomogeneous field configurations \cite{jangir2026dynamically, orthaber2011momentum, fey2012momentum, nuriman2012enhanced, otto2015, schneider2016dynamically, aleksandrov2018dynamically, kohlfurst2021dynamically, li2021enhanced}. 
	
	Frequency modulation provides another route to enhancing pair production: the introduction of a suitable chirp can significantly increase the pair yield while inducing pronounced modifications in the momentum spectrum. Such effects have been studied in both spatially homogeneous time-dependent fields \cite{dumlu2010schwinger, gong2020electron, olugh2019pair, jangir2026dynamically} and spatially inhomogeneous oscillating backgrounds \cite{ababekri2020chirp, li2021enhanced, bake2025vacuum, osman2023efficient}. In addition, temporal asymmetry of the electric-field pulse has been shown to provide a substantial enhancement of pair production \cite{chen2024asymmetric, jangir2026carrier, oluk2014electron, olugh2020asymmetric}. In asymmetric fields, the unequal temporal evolution of the field on the two sides of the pulse modifies the interference between different pair-production events and alters the associated tunneling dynamics, which can lead to significant changes in both the total yield and the momentum distribution.
	
	Beyond linearly polarized electric fields, considerable attention has also been devoted to vacuum pair production in fields with more general, in particular nonlinear, polarization structures. In recent years, spin-, chirality-, and helicity-dependent signatures have emerged as important aspects of nonperturbative pair production. Studies of polarized and rotating electromagnetic backgrounds have shown that the external-field configuration influences the spin properties of the produced $e^+e^-$ pairs \cite{wollert2015spin, blinne2016comparison, li2017momentum, olugh2019pair, kohlfurst2019spin, jangir2026dynamically, olugh2019pair, olugh2020asymmetric, li2015effects, blinne2014rotating}. Using the DHW formalism, these effects have been investigated through, among others, spiral structures in momentum space, chirality-dependent phenomena, and spin asymmetries \cite{hu2023momentum, yu2023evolution, jiang2026study, aleksandrov2024pair, majczak2024scattering}. For example, Hu \textit{et al.} \cite{hu2024spin} investigated circularly polarized fields with different frequencies and relative time delays, and demonstrated the emergence of spiral structures in both spin-up and spin-down momentum distributions. They further showed that the degree of spin asymmetry can be substantially modified by varying the time delay between the two fields. In a related study, Chen \textit{et al.} \cite{chen2025spin} considered two counter-rotating circularly polarized electric fields and found that the spin polarization of the produced pairs is highly sensitive to the pulse shape, chirp, and temporal delay between the pulses. These results highlight the strong dependence of spin-related observables on the detailed structure of the driving field.
	
	Recent investigations have explored helicity effects in a variety of contexts, including the dynamics of twisted particles \cite{ivanov2016quantum, sherwin2020effect, ivanov2020twisted}, condensed-matter systems \cite{quessab2018helicity, zhang2019energy}, and particle-physics \cite{agarwal2021two, chawdhry2021two}. In the context of strong-field pair production, Aleksandrov \textit{et al.} \cite{aleksandrov2024kinetic} recently revisited the quantum kinetic equations (QKEs) for $e^+e^-$ production in spatially homogeneous, time-dependent electric fields with arbitrary polarization. They demonstrated the equivalence of the corrected QKE formulation with the DHW formalism and, within this framework, derived both spin-resolved and helicity-resolved momentum distribution functions. Subsequently, Jiang \textit{et al.} \cite{jiang2026study} showed that the helicity-resolved distribution can be understood as a particular case of the more general spin-resolved distribution. Specifically, the helicity-resolved distribution is recovered by choosing the spin-quantization direction to be parallel to the momentum of the produced particle. This connection provides a direct link between the more general spin-resolved description and helicity as a momentum-dependent spin observable. In a recent study, Baksheev \textit{et al.} \cite{baksheev2026helicity} investigated helicity-resolved observables in dynamically assisted electric fields within the dipole approximation. Building upon these developments, it is pertinent to investigate how the helicity signatures of the produced pairs are influenced by the temporal characteristics of the driving field, with particular emphasis on the pulse shape and frequency chirping.
	
	In this work, we employ the QKE framework to investigate helicity signatures in vacuum pair production induced by spatially homogeneous, time-dependent rotating electric fields. We consider two distinct classes of field configurations: (i) temporally asymmetric electric fields with frequency chirping, and (ii) two-color dynamically assisted electric fields in which the effects of frequency chirping are examined in different configurations. For the latter, we consider three distinct scenarios: chirping applied exclusively to the strong-field component, chirping applied exclusively to the weak-field component, and simultaneous chirping of both components. For completeness, we also examine the strong- and weak-field components of the dynamically assisted configuration separately, focusing on how frequency modulation modifies their respective helicity structures.

	The paper is organized as follows. In Sec. \ref{sec:theoretical_formalism}, we briefly introduce the QKE framework employed in our numerical analysis and define the helicity-resolved observables considered throughout this work. Section \ref{sec:asym_fields} presents our results for temporally asymmetric electric fields, while Sec. \ref{sec:DASE_fields} is devoted to the individual strong- and weak-field components and their combined two-color dynamically assisted configuration. Finally, our main findings are summarized and the conclusions are presented in Sec. \ref{sec:conclusion}.
	\section{Theoretical formalism - Quantum Kinetic Equations}\label{sec:theoretical_formalism}
	The analysis presented in this work is based primarily on the QKEs, which provide a nonperturbative framework for studying $e^+e^-$ pair production from the vacuum in the presence of strong background fields with arbitrary polarization. A detailed derivation of the QKE formalism can be found in Refs. \cite{aleksandrov2024kinetic, aleksandrov2026vacuum}. Here, we briefly review the essential equations and definitions relevant to the numerical calculations presented in this work. Throughout this work, we employ natural units, $\hbar=c=1$, and take $e=1$ as the positive magnitude of the elementary charge. Accordingly, we adopt the convention $\textbf{p}(t) = \textbf{k} - e\textbf{A}(t)$ where $\textbf{k}$ and $\textbf{p}(t)$ denote the canonical and kinetic momentum, respectively.
	
	For a spatially homogeneous, time-dependent arbitrarily polarized electric field $\textbf{E}(t)$, the $e^+e^-$ pair-production dynamics can be described by a system of coupled ordinary differential equations for ten unknown functions $f$, $\textbf{f}$, $\textbf{u}$, and $\textbf{v}$:
	\begin{align}
	\dot{f} &= -2\bm{\mu}_2 \cdot \textbf{u},
	\label{eq:QKE1}\\
	\dot{\textbf{f}} &= 2(\bm{\mu}_1 \times \textbf{f})
	-2(\bm{\mu}_2 \times \textbf{v}),
	\label{eq:QKE2}\\
	\dot{\textbf{u}} &= 2(\bm{\mu}_1 \times \textbf{u})
	+\bm{\mu}_2(2f-1)
	+2\omega\textbf{v},
	\label{eq:QKE3}\\
	\dot{\textbf{v}} &= 2(\bm{\mu}_1 \times \textbf{v})
	-2(\bm{\mu}_2 \times \textbf{f})
	-2\omega\textbf{u}.
	\label{eq:QKE4}
	\end{align}
	Here, the time-dependent quantities $\bm{\mu}_1$ and $\bm{\mu}_2$ contain the coupling between the quantum kinetic variables and the external electric field. They are defined as
	\begin{align}
		\bm{\mu}_1(\textbf{k},t)
		&=
		\frac{e}{2\Omega(\Omega+m)}
		\left[\textbf{p}\times\textbf{E}(t)\right],
		\label{eq:QKE_auxilary1}\\
		\bm{\mu}_2(\textbf{k},t)
		&=
		\frac{e}{2\Omega^2(\Omega+m)}
		\left\{
		[\textbf{p}\cdot\textbf{E}(t)]\textbf{p}
		-\Omega(\Omega+m)\textbf{E}(t)
		\right\}.
		\label{eq:QKE_auxilary2}
	\end{align}
	The momentum dependence of the QKE system is expressed in terms of the canonical momentum $\textbf{k}$. We work in the temporal gauge, with the four-potential given by $A^\mu(t)=(0,\mathbf{A}(t))$. The vector potential is related to the prescribed external electric field through $\mathbf{E}(t)=-\dot{\mathbf{A}}(t)$, which specifies the classical electromagnetic background driving the pair-production process. The instantaneous energy of the particle (of mass $m$) is $\Omega(t)= \sqrt{m^2+\textbf{p}(t)^{2}}$. Thus, for each fixed canonical momentum $\textbf{k}$, the QKE system \eqref{eq:QKE1}--\eqref{eq:QKE4} is evolved in time, with the external field entering through $\bm{\mu}_1$, $\bm{\mu}_2$, and the kinetic momentum $\textbf{p}(t)$.
	
	To describe pair production from the vacuum, the QKE functions are initialized in the vacuum state at an initial time $t=t_{\mathrm{in}}\rightarrow-\infty$, with all QKE functions set to zero. The equations are then evolved to a sufficiently large final time $t=t_{\mathrm{out}}\rightarrow+\infty$, chosen such that the external electric field has effectively vanished at the integration boundaries. In this asymptotic region, the kinetic momentum becomes time independent. For a gauge in which $\textbf{A}(t_{\mathrm{out}})=0$, one has $\textbf{p}(t_{\mathrm{out}})=\textbf{k}$. Consequently, the canonical momentum $\textbf{k}$ can be identified with the asymptotic kinetic momentum of the produced particles. The QKE functions evaluated at $t=t_{\mathrm{out}}$ can then be used to obtain the observable momentum distributions.

	In particular, the helicity-resolved momentum distributions of the produced electrons are given by \cite{aleksandrov2024kinetic}
	\begin{equation}
		f^{(e^-)}_{\text{L/R}}(\textbf{k},t)=f(\textbf{k},t)\mp\frac{\textbf{p}\cdot\textbf{f}(\textbf{k},t)}{|\textbf{p}|}.
		\label{eq:helicity-resolved_electron_MD}
	\end{equation}
	Here, $L$ and $R$ denote the left- and right-handed helicity states, corresponding to negative and positive helicity, respectively. Evaluating the above distributions at the asymptotic time $t=t_\text{out}$, one obtains 
	\begin{equation}
		\frac{(2\pi)^3}{V}
		\frac{dN_{L/R}^{(e^-)}(\textbf{k})}{d^3\textbf{k}}
		=
		f^{(e^-)}_\text{L/R}(\textbf{k})
		\equiv
		f^{(e^-)}_\text{L/R}(\textbf{k},t_{\text{out}}),
		\label{eq:Production_rate}
	\end{equation}
	where $V$ denotes the normalization volume. The helicity-resolved distributions satisfy $f^{(e^-)}_\text{L/R}(\textbf{k})\leq 1$, consistent with the Pauli exclusion principle \cite{aleksandrov2024kinetic}. The helicity-summed, equivalently spin-summed, electron momentum distribution is then given by
	\begin{equation}\label{eq:helicity_summed_electron_spectrum}
		f^{(e^-)}(\textbf{k}) = f^{(e^-)}_\text{L}(\textbf{k}) + f^{(e^-)}_\text{R}(\textbf{k}) = 2f(\textbf{k}).
	\end{equation}
	The corresponding positron distributions can be obtained from
	\begin{equation}\label{eq:helicity_resolved_positron_spectrum}
		f^{(e^+)}_\text{L/R}(\textbf{k},t) = f^{(e^-)}_\text{R/L}(-\textbf{k},t).
	\end{equation}
	In the following sections, we numerically solve the QKE system \eqref{eq:QKE1}--\eqref{eq:QKE4} and analyze both the helicity-summed and helicity-resolved momentum distributions, $f^{(e^-)}(\textbf{k})$ and $f^{(e^-)}_{\mathrm{L/R}}(\textbf{k})$, respectively.
	\subsection{Numerical implementation}\label{subsec:numerical_implementation}
	The QKE system \eqref{eq:QKE1}--\eqref{eq:QKE4} is integrated using the adaptive Dormand--Prince $8(5,3)$ method over the time interval $t\in[-\eta,\eta]$. Here, $\eta>0$ is chosen sufficiently large such that the external electric field is effectively zero at both integration boundaries. The numerical implementation is benchmarked against Refs.~\cite{aleksandrov2024pair, chen2024asymmetric}, with excellent agreement with the published results. For the calculations presented in this work, a uniform momentum grid with spacing $\Delta k=0.01m$ is employed to obtain sufficiently fine resolution of the momentum distributions.
	\subsection{Helicity asymmetry and choice of momentum-space plane}\label{subsec:helicity_asymmetry}
	In sections \ref{sec:asym_fields} and \ref{sec:DASE_fields}, we consider two different temporal forms of the electric field, both of which are confined to the $(x,y)$ plane. In addition to the helicity-resolved momentum distributions introduced above (cf. Eq. \eqref{eq:helicity-resolved_electron_MD}), we use the \textit{normalized helicity asymmetry} (or simply \textit{helicity asymmetry})
	\begin{equation}\label{eq:helicity_asymmetry}
		a^{(e^-)}(\mathbf{k})
		=
		\frac{
			f^{(e^-)}_{\mathrm{L}}(\mathbf{k})
			-
			f^{(e^-)}_{\mathrm{R}}(\mathbf{k})
		}{
			\max\left[
			f^{(e^-)}_{\mathrm{L}}(\mathbf{k})
			+
			f^{(e^-)}_{\mathrm{R}}(\mathbf{k})
			\right]
		},
	\end{equation}
	to quantify the difference between the two helicity states. Using Eq.~\eqref{eq:helicity-resolved_electron_MD}, the helicity asymmetry can be written as
	\begin{equation}
		a^{(e^-)}(\mathbf{k})
		=
		-\frac{2\,\mathbf{p}\cdot\mathbf{f}(\mathbf{k})}
		{|\mathbf{p}|\,
			\max\left[
			f^{(e^-)}_{\mathrm{L}}(\mathbf{k})+
			f^{(e^-)}_{\mathrm{R}}(\mathbf{k})
			\right]}.
		\label{eq:helicity_asymmetry_QKE}
	\end{equation}
	Thus, the helicity asymmetry is directly determined by the component of $\mathbf{f}$ along the kinetic momentum $\mathbf{p}$.
	
	For the present planar electric field, the helicity asymmetry exhibits a strong dependence on the momentum-space plane in which it is evaluated. In the $(k_x,k_y)$ plane, corresponding to $k_z = 0$, both the kinetic momentum and the electric field lie in the $(x,y)$ plane. Consequently, $\bm{\mu}_1$ is directed along the $z$ axis, while $\bm{\mu}_2$ lies in the $(x,y)$ plane. With vacuum initial conditions, the QKE system \eqref{eq:QKE1}--\eqref{eq:QKE4} then preserves the structure $\mathbf{f}\parallel\hat{z}$, yielding $\textbf{p}\cdot\mathbf{f} = 0$ and hence $f_L = f_R$. In contrast, for $(k_x,k_z)$ and ($k_y,k_z$) plane, corresponding to $k_y = 0$ and $k_x = 0$, the kinetic momentum has a nonzero $z$-component, and $\bm{\mu}_1$ and $\bm{\mu}_2$ generally acquire components outside the planar configuration. The QKE evolution can therefore generate a nonzero component of $\mathbf{f}$ parallel to $\mathbf{p}$, resulting in an appreciable helicity asymmetry. We therefore focus on helicity-resolved momentum distributions in the $(k_x,k_z)$ plane with $k_y=0$, where a nonvanishing helicity asymmetry can arise. In contrast, most previous studies have focused on momentum spectra in the $(k_x,k_y)$ polarization plane \cite{blinne2014rotating, blinne2016comparison, li2015effects, li2017momentum, jangir2026dynamically, olugh2019pair, olugh2020asymmetric, hu2024spin, hu2023momentum}. The comparatively less explored $(k_x,k_z)$ plane is therefore investigated here.
	
 	In what follows, we introduce two temporal forms of the electric field considered in this work and investigate how their temporal structure and frequency chirping affect the helicity-resolved momentum distributions and the corresponding helicity asymmetry.
	\section{Temporally asymmetric electric fields}\label{sec:asym_fields}
	We first consider a temporally asymmetric, circularly polarized electric field with linear-frequency chirp. The field is constructed by assigning different temporal envelope widths to the two sides of the pulse,
	\begin{equation}\label{eq:asymmetric_field}
		\begin{split}
			\textbf{E}(t)  = &\frac{E_0}{\sqrt{2}}\left[\text{exp}\left(-\frac{t^2}{2\tau_1^2}\right)\Theta(-t) + \text{exp}\left(-\frac{t^2}{2\tau_2^2}\right)\Theta(t)\right]\\
			& \times \begin{pmatrix}
				\cos(\omega t + bt^2)\\
				\sin(\omega t + bt^2)\\
				0
			\end{pmatrix}.
		\end{split}
	\end{equation}
	Here, $E_0$ is the overall field amplitude, $\omega$ is the original carrier frequency, and $b$ is the chirp parameter. The two Gaussian envelopes are characterized by the temporal widths $\tau_1$ and $\tau_2$ for the $t<0$ and $t>0$ portions of the pulse, respectively. The Heaviside functions $\Theta(-t)$ and $\Theta(t)$ restrict the corresponding envelopes to their respective time domains. For $\tau_1=\tau_2$, the temporal envelope is symmetric about $t=0$. A temporal asymmetry is introduced by choosing $\tau_1\neq\tau_2$. We characterize the degree of asymmetry by defining the \textit{pulse asymmetry parameter} $\beta = \tau_2/\tau_1$. The falling edge ($t>0$) of the pulse is compressed for $\beta<1$ and elongated for $\beta>1$. In the following calculations, we fix the field amplitude, carrier frequency, and width of the $t<0$ portion to
	\begin{equation}\label{eq:asymmetric_field_parameters}
		E_0 = 0.1\sqrt{2}E_\text{cr},\quad \omega=0.6\,m,\quad \tau_1=10/m,
	\end{equation}
	while $\tau_2$ is varied through the asymmetry parameter $\beta$. The chirp parameter $b$ is varied independently over the range $0 \leq b \leq 0.9\omega/\tau_1$, corresponding to the \textit{normal-chirp regime} \cite{olugh2019pair, jangir2026dynamically}, to investigate the combined effects of temporal asymmetry and frequency chirping. Pair production can proceed through the Schwinger or perturbative multiphoton mechanism, characterized by the Keldysh parameter $\gamma=m\omega/eE$ \cite{keldysh1965ionization}, with $\gamma\ll1$ and $\gamma\gg1$ corresponding to the tunneling and multiphoton regimes, respectively. For the parameters in Eq. \eqref{eq:asymmetric_field_parameters}, the unchirped field in Eq. \eqref{eq:asymmetric_field} lies in the intermediate regime, $\gamma\sim\mathcal{O}(1)$. For a chirped field, the effective Keldysh parameter is obtained by replacing $\omega$ with $\omega_{\mathrm{eff}}=\omega+bt$.
		\begin{figure*}[tbh]
		\centering
		\includegraphics[width=\textwidth]{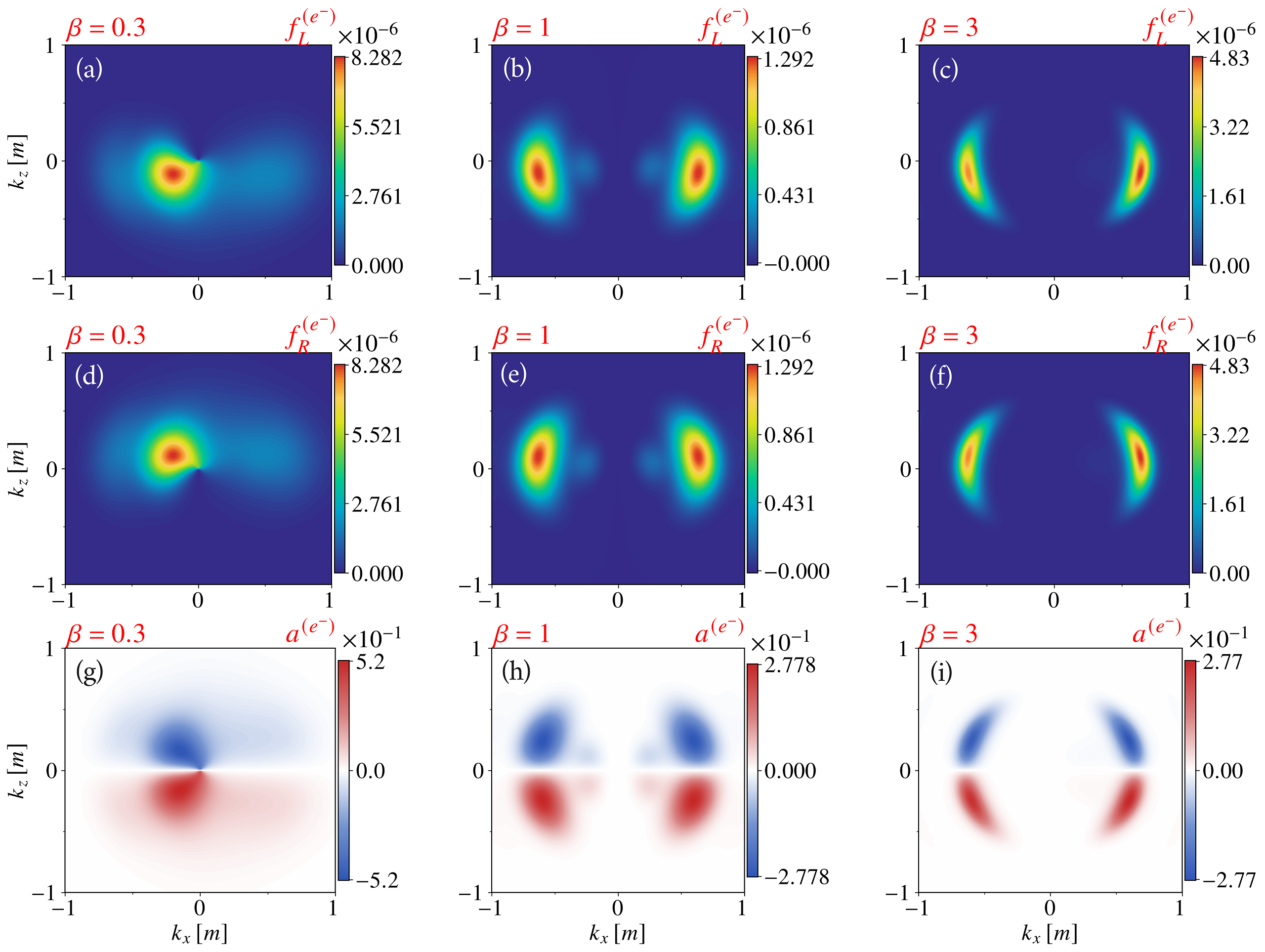}
		\caption{Helicity-resolved momentum distributions of the electrons produced in electric field \eqref{eq:asymmetric_field}, at $k_y=0$ for chirp-free case, $b=0$. Panels (a)--(c) show the left-helicity component $f_L^{(e^-)}$, panels (d)--(f) show the right-helicity component $f_R^{(e^-)}$, and panels (g)--(i) show the corresponding helicity asymmetry $a^{(e^-)}$. From left to right, the columns (a)--(g), (b)--(h) and (c)--(i) correspond to the pulse asymmetry parameter $\beta=0.3, 1,$ and $3$, respectively. The remaining field parameters are the same as those given in Eq. \eqref{eq:asymmetric_field_parameters}.}
		\label{fig:asym_MD_(kx,kz)_alpha_0}
	\end{figure*}
	
	The electric field in Eq.~\eqref{eq:asymmetric_field} is considered within the \textit{dipole approximation}, representing the spatially homogeneous limit of a realistic space-time-dependent field. This idealized setup allows us to isolate the effects studied here; limitations of the approximation are discussed in Refs.~\cite{lv2018, aleksandrov2017momentum}. A fully space-time-dependent treatment is left for future work.
	
	The temporally asymmetric electric field defined in Eq.~\eqref{eq:asymmetric_field} has been extensively investigated in previous studies, particularly with regard to its helicity-summed momentum distributions \cite{chen2024asymmetric, olugh2020asymmetric, jangir2026carrier}. In the present work, we revisit this field configuration from a helicity-resolved perspective and analyze the corresponding momentum distributions and helicity asymmetry of the produced particles.
	\subsection{Chirp-free fields $b = 0$}\label{subsec:asym_field_chirp_free}
	\begin{figure*}[tbh]
		\centering
		\includegraphics[width=\textwidth]{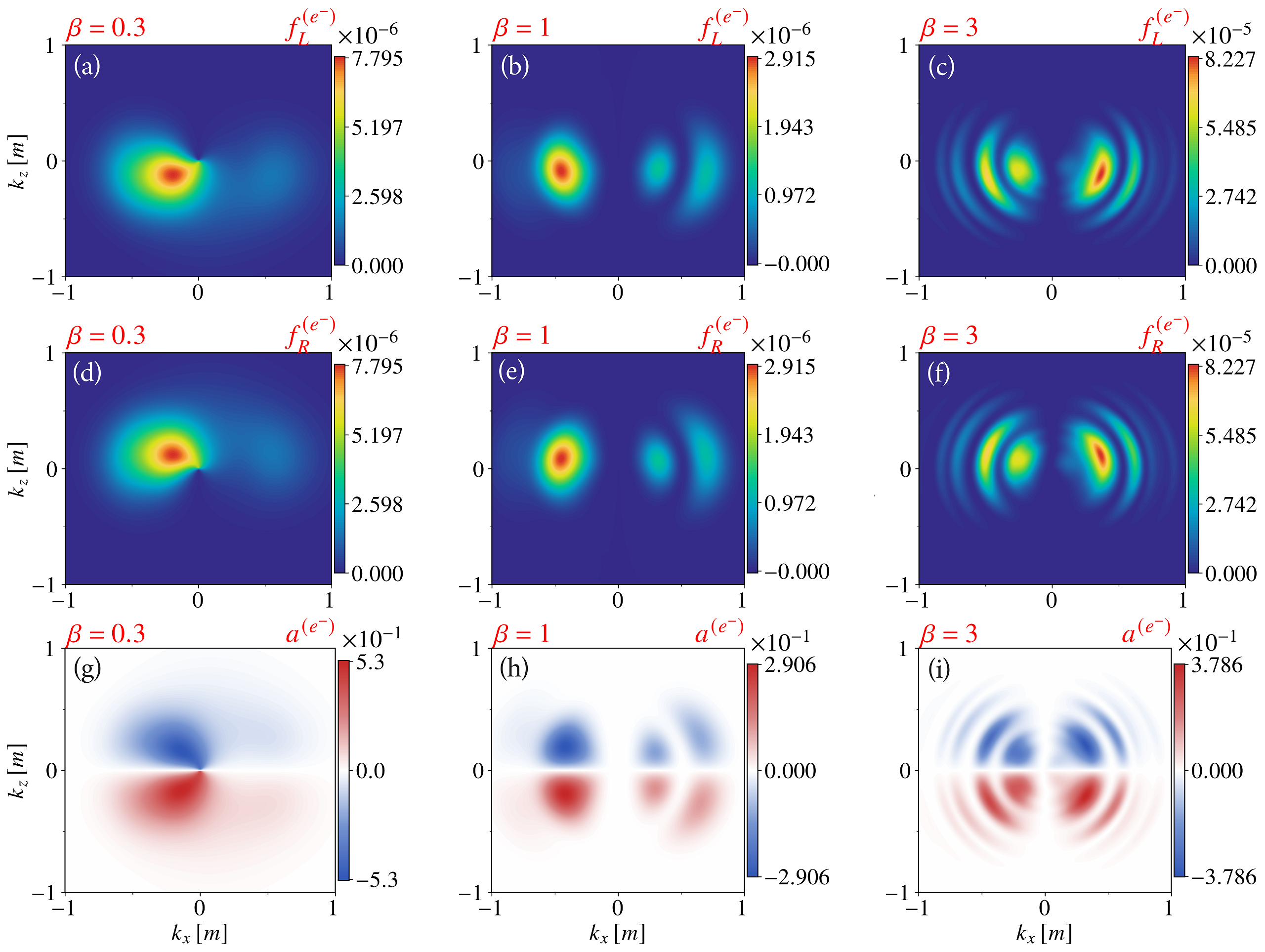}
		\caption{Same as Fig. \ref{fig:asym_MD_(kx,kz)_alpha_0} except for $b=0.1\omega/\tau_1 = 0.006\,m^2$.}
		\label{fig:asym_MD_(kx,kz)_alpha_0.1}
	\end{figure*}
	
	\begin{figure*}[tbh]
		\centering
		\includegraphics[width=\textwidth]{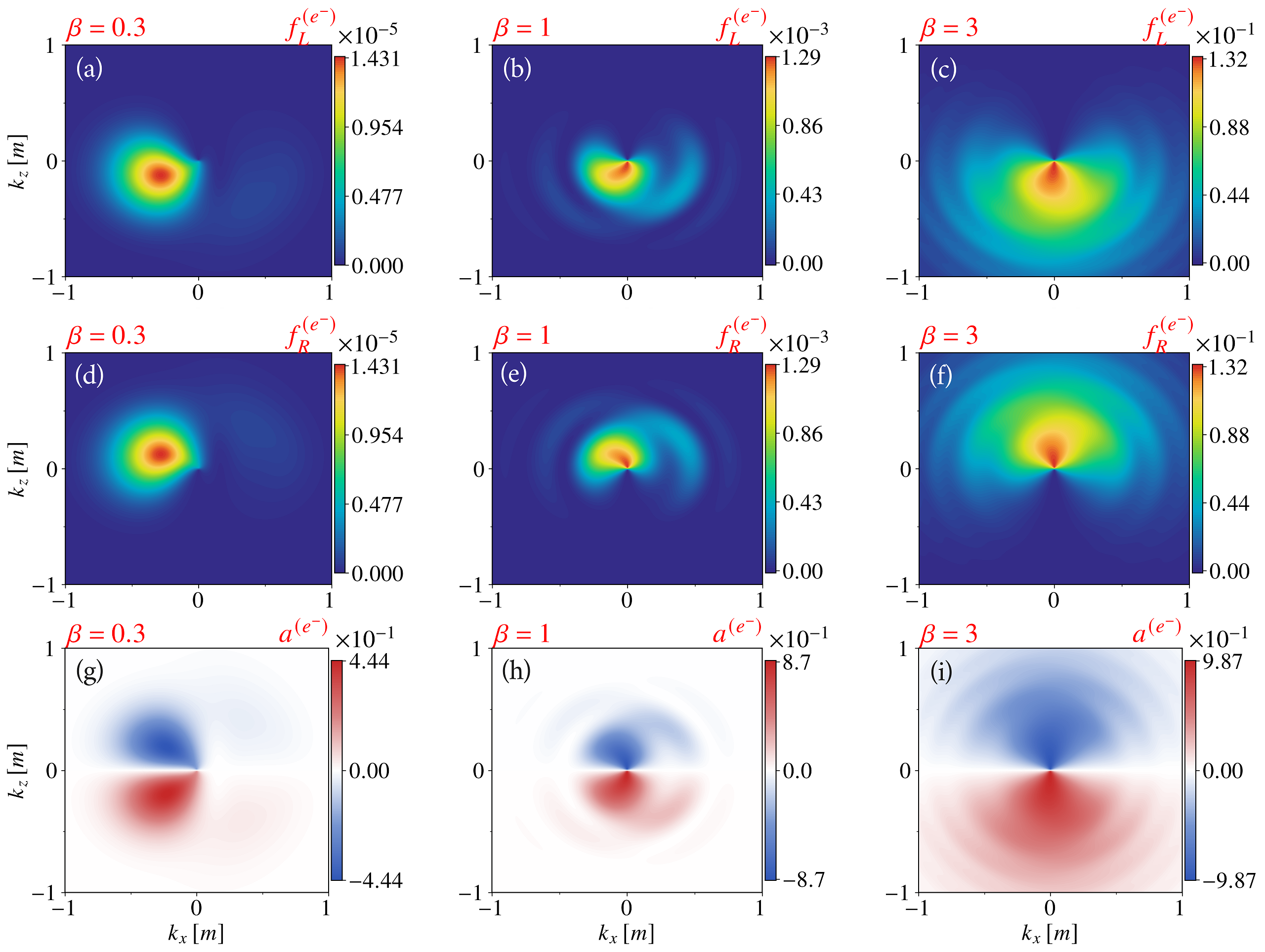}
		\caption{Same as Fig. \ref{fig:asym_MD_(kx,kz)_alpha_0} except for $b=0.5\omega/\tau_1 = 0.03\,m^2$.}
		\label{fig:asym_MD_(kx,kz)_alpha_0.5}
	\end{figure*}
	Figure~\ref{fig:asym_MD_(kx,kz)_alpha_0} presents the helicity-resolved momentum distributions and the corresponding helicity asymmetries of the produced electrons in the $(k_x$,$k_z)$ plane at $k_y=0$ for the chirp-free case ($b = 0$), and for three representative values of the pulse asymmetry parameter, $\beta=0.3$, $1$, and $3$. For each value of $\beta$, the helicity asymmetry, defined in Eq. \eqref{eq:helicity_asymmetry}, is normalized with respect to the maximum of the helicity-summed momentum distribution over the entire $(k_x,k_z)$ plane for that particular value of $\beta$. Thus, the values of $a^{(e^-)}$ shown in Figs. \ref{fig:asym_MD_(kx,kz)_alpha_0}(g)--\ref{fig:asym_MD_(kx,kz)_alpha_0}(i) represent the helicity difference relative to the maximum helicity-summed pair-production yield for the corresponding pulse configuration. 
	
	The helicity-resolved distributions exhibit characteristic structures, whose location and overall peak magnitude are modified by the temporal asymmetry of the pulse. Although the left- and right-helicity components have similar overall momentum support, their intensities are distributed differently across the $(k_x,k_z)$ plane, revealing a clear momentum-dependent separation between the two helicity states. In particular, the left-helicity distribution is predominantly concentrated in the negative $k_z$ region, whereas the right-helicity component is enhanced toward positive $k_z$ region.  The magnitudes of the peak positions along the $z$ direction remain approximately unchanged with the pulse asymmetry, with the left- and right-helicity distributions peaking at $k_z\simeq-0.11m$ and $k_z\simeq+0.11m$, respectively. The overall magnitude of the momentum distributions, however, is strongly sensitive to the temporal asymmetry. Among the three cases considered, the largest peak yield occurs for $\beta=0.3$. The enhanced production for $\beta<1$ can be attributed to the steep switch-off of the electric field \cite{aleksandrov2025switchoff}, which is known to enhance pair production in temporally asymmetric pulse configurations \cite{jangir2026carrier}.
	
	The momentum-dependent separation of the two helicity components is further quantified by the helicity asymmetry. A clear sign reversal
	occurs across the $k_z=0$ axis: $a^{(e^-)}>0$ predominantly in the negative-$k_z$ region, indicating a relative enhancement of
	left-helicity electrons, whereas $a^{(e^-)}<0$ for positive $k_z$, indicating a relative enhancement of right-helicity electrons. For $\beta=0.3$, the maximum magnitude of the asymmetry reaches approximately $52\%$, while the corresponding maximum magnitude is about $28\%$ for the other two pulse configurations. 
	
	The odd dependence of the helicity asymmetry on $k_z$, is consistent with the discussion made in Sec.~\ref{subsec:helicity_asymmetry}. For the present planar electric field, $A_z=0$ and hence $p_z=k_z$. Since the helicity difference is determined by $f^{(e^-)}_{\mathrm L}-f^{(e^-)}_{\mathrm R}=-2{\mathbf{p}\cdot\mathbf{f}}/{|\mathbf{p}|}$, the change $k_z\rightarrow-k_z$ reverses the helicity difference. 
	
	Overall, even in the absence of frequency chirping, the rotating electric field produces a clear momentum-helicity correlation, with left-helicity electrons preferentially associated with negative $k_z$ and right-helicity electrons with positive $k_z$. The temporal asymmetry primarily modifies the magnitude and detailed momentum-space structure of this correlation, while the characteristic odd-in-$k_z$ helicity asymmetry follows from the QKE dynamics and the three-dimensional momentum geometry of the planar rotating field.
	\subsection{small chirp parameter $b = 0.1\omega/\tau_1$}\label{subsec:asym_field_chirp_small}
	\begin{figure*}[tbh]
		\centering
		\includegraphics[width=\textwidth]{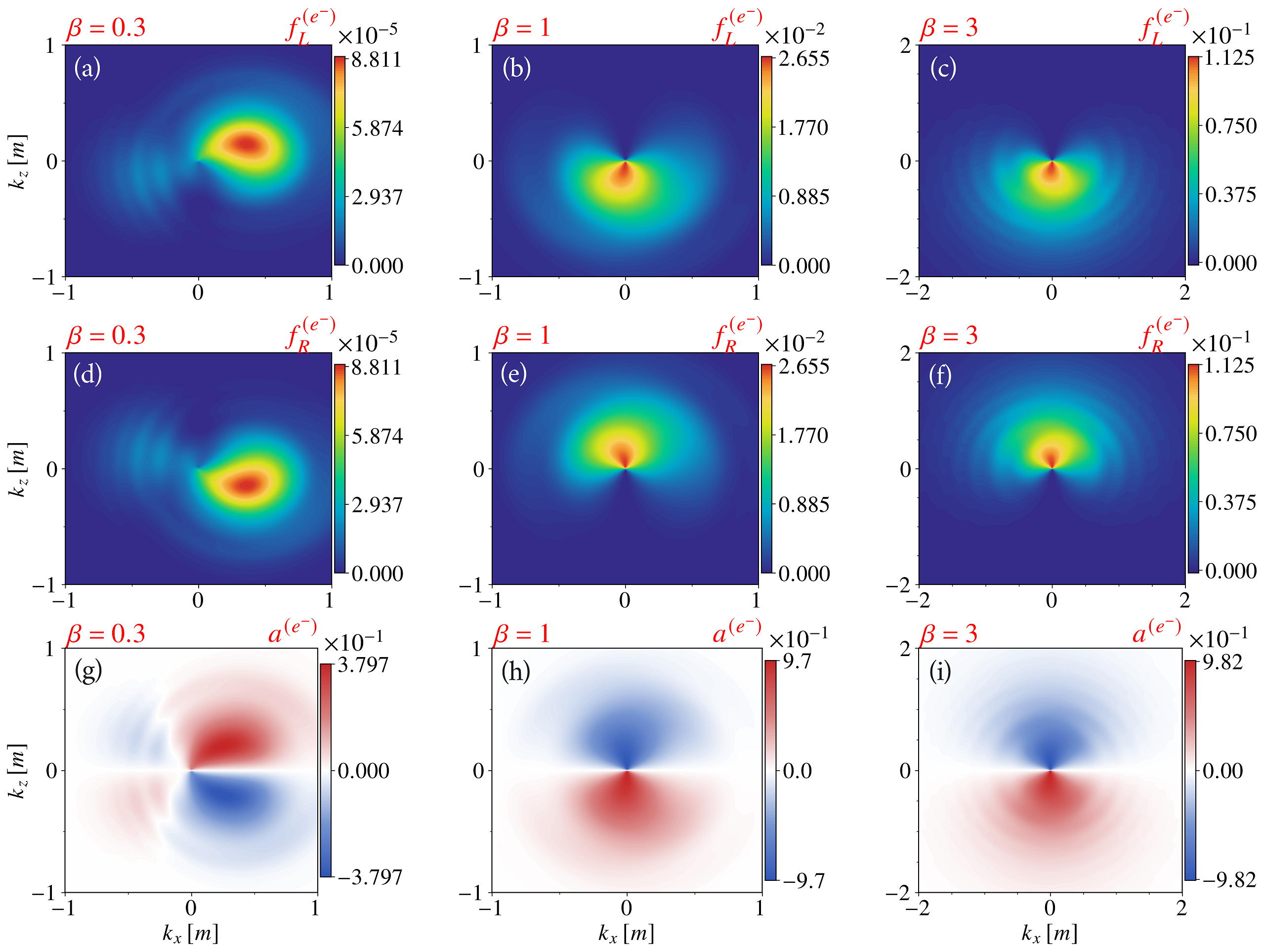}
		\caption{Same as Fig. \ref{fig:asym_MD_(kx,kz)_alpha_0} except for $b=0.9\omega/\tau_1 = 0.054\,m^2$.}
		\label{fig:asym_MD_(kx,kz)_alpha_0.9}
	\end{figure*}
	For a weak linear chirp, $b_1=0.1\omega/\tau_1=0.006\,m^2$, the corresponding helicity-resolved momentum distributions and helicity asymmetries are shown in Fig.~\ref{fig:asym_MD_(kx,kz)_alpha_0.1}. The introduction of the chirp modifies both the shape and magnitude of the
	momentum spectra, with the effect becoming particularly pronounced for $\beta = 3$. For $\beta=0.3$ and $\beta=1$, the
	distributions remain relatively smooth and localized around the dominant momentum structures, whereas for $\beta=3$ the spectra develop an oscillatory, multi-lobed structure extending over a wider region of the $(k_x,k_z)$ plane. 
	
	While the peak values of the helicity-resolved momentum distributions is of the same order as in the chirp-free case for $\beta=0.3$ and $\beta=1$, a substantial enhancement occurs for $\beta=3$, where the peak reaches the order of $10^{-5}$. The dominant momentum structures remain centered at finite $k_x$ and occur at opposite values of $k_z$ for the two helicity components. For all the $\beta$ values shown, the principal peaks remain located close to $k_z\simeq\mp0.1m$ for the left- and right-helicity components, respectively. The oscillatory structures, in $\beta = 3$, can be associated with the enhanced frequency content of the chirped field and the extended duration of the falling edge, which allow the field to undergo more oscillation cycles and provide additional opportunities for multiphoton absorption \cite{blinne2016PhD} during pair production. Consequently, the particles can experience different amounts of acceleration at different stages of the pulse, giving rise to the observed oscillatory structures in the momentum spectrum \cite{jangir2026carrier, chen2024asymmetric}.

	The corresponding helicity asymmetry retains the characteristic momentum-dependent separation between the two helicity states. The
	maximum magnitudes of $a^{(e^-)}$ are corresponding to approximately $53\%$, $29\%$, and $38\%$ of the maximum helicity-summed momentum distribution for $\beta = 0.3, 1$ and $3$, respectively. The asymmetry remains predominantly positive in the negative-$k_z$ region and negative in the positive-$k_z$ region. For $\beta=3$, the helicity-asymmetry distribution develops alternating regions of enhanced and suppressed asymmetry, which follow the oscillatory structures observed in the corresponding helicity-resolved momentum distributions.
	\subsection{Intermediate chirp parameter $b = 0.5\omega/\tau_1$}\label{subsec:asym_field_chirp_intermediate}
	For the intermediate chirp, $b_1=0.5\omega/\tau_1=0.03\,m^2$, the helicity-resolved momentum distributions and the corresponding helicity asymmetry are shown in Fig.~\ref{fig:asym_MD_(kx,kz)_alpha_0.5}. The momentum spectra become progressively broader and more structured with
	increasing pulse asymmetry, with the appearance of additional oscillatory features for $\beta=1$ and $\beta=3$. The most pronounced modification
	occurs for $\beta=3$, where the distribution develops an extended multi-lobed structure over the $(k_x,k_z)$ plane. The peak values of the helicity-resolved momentum distributions indicate a very strong enhancement of the pair-production yield with increasing $\beta$ at this chirp strength. The dominant helicity-resolved structures remain centered around finite $k_x$ and opposite $k_z$ regions for the two helicities.
	
	The helicity asymmetry retains the characteristic separation of the two helicity states, with positive asymmetry predominantly in the
	negative-$k_z$ region and negative asymmetry in the positive-$k_z$ region. The helicity asymmetry is approximately $44\%$, $87\%$, and $99\%$ relative to the maximum helicity-summed momentum distribution for $\beta = 0.3, 1$ and $3$, respectively. Thus, at this intermediate chirp strength, the helicity imbalance becomes particularly pronounced for the temporally asymmetric field with large falling edge pulse, reaching nearly $100\%$ of the chosen normalization for $\beta=3$.
	\subsection{Large chirp parameter $b = 0.9\omega/\tau_1$}\label{subsec:asym_field_chirp_large}
	For the large chirp, $b_1=0.9\omega/\tau_1=0.054\,m^2$, the helicity-resolved momentum distributions and the corresponding helicity
	asymmetry are shown in Fig.~\ref{fig:asym_MD_(kx,kz)_alpha_0.9}. The momentum spectra undergo a substantial reshaping compared with the
	smaller-chirp cases. For $\beta=0.3$, the distributions remain localized in a relatively narrow momentum region but develop additional oscillatory structures on the opposite side of the dominant peak. For $\beta=1$ and $3$, the spectra become considerably broader and smoother,
	with a pronounced central structure extending over a larger momentum region. The broadening is particularly evident for $\beta=3$, where the
	momentum support extends to approximately $|k_x|,|k_z|\sim 2m$.
	
	The peak values of the helicity-resolved momentum distributions increase dramatically with increasing $\beta$, changing from the $10^{-5}$ level for $\beta=0.3$ to the $10^{-2}$ and $10^{-1}$ levels for $\beta=1$ and $3$, respectively. The dominant left- and right-helicity structures remain displaced toward opposite $k_z$ regions, although their detailed locations and shapes are substantially modified by the chirp. Unlike the previous chirping cases considered, however, the $\beta=0.3$ case under strong chirping [Figs. \ref{fig:asym_MD_(kx,kz)_alpha_0.9}(a)  and \ref{fig:asym_MD_(kx,kz)_alpha_0.9}(d)] exhibits an opposite helicity signature: the left-helicity contribution is predominantly localized in the positive-$k_z$ region, while the right-helicity contribution dominates in the negative-$k_z$ region.
	
	For $\beta=0.3$, $1$, and $3$, the helicity asymmetry reaches approximately $38\%$, $97\%$, and $98\%$, respectively, relative to the maximum helicity-summed momentum distribution for the corresponding pulse configurations. For $\beta=1$ and $3$, the asymmetry retains the characteristic sign structure in which the negative-$k_z$ region is dominated by left-helicity electrons and the positive-$k_z$ region by right-helicity electrons. For the strongly compressed pulse, $\beta=0.3$, however, the chirp-induced spectral reshaping is accompanied by a reversal of this dominant helicity ordering, with the positive-$k_z$ region exhibiting positive asymmetry and the negative-$k_z$ region negative asymmetry. This indicates that a sufficiently strong chirp can modify not only the magnitude but also the momentum-helicity correlation for strongly
	asymmetric pulses.
	
	Taken together, the results in Table \ref{tab:max_helicity_asymmetry_asymmetric} demonstrate a strong dependence of the helicity asymmetry on both the chirp strength and the pulse asymmetry parameter $\beta$. For the nearly symmetric and temporally extended cases, $\beta=1$ and $\beta=3$, the asymmetry increases strongly with chirp, reaching $97\%$ and $98.2\%$ at the largest chirp strength, respectively. In contrast, for $\beta=0.3$, the asymmetry decreases from $53\%$ at weak chirp to $37.97\%$ at the largest chirp. Thus, increasing chirp strongly enhances the helicity asymmetry for $\beta\geq1$, whereas the strongly compressed pulse exhibits a qualitatively different response, with the asymmetry being suppressed at large chirp.
	\begin{table}[t]
		\centering
		\caption{Maximum helicity asymmetry, expressed as a percentage, for the temporally asymmetric electric field \eqref{eq:asymmetric_field} with different pulse asymmetry parameters $\beta$ and chirp strengths.}
		\label{tab:max_helicity_asymmetry_asymmetric}
		\renewcommand{\arraystretch}{1.35}
		\begin{ruledtabular}
			\begin{tabular*}{\textwidth}{@{\extracolsep{\fill}}cccc}
				Chirp strength
				& $\beta=0.3$ & $\beta=1$ & $\beta=3$ \\
				\hline
				$b=0$
				& $52.00\%$ & $27.78\%$ & $27.70\%$ \\
				
				$b=0.1\omega/\tau_1$
				& $53.00\%$ & $29.06\%$ & $37.86\%$ \\
				
				$b=0.5\omega/\tau_1$
				& $44.40\%$ & $87.00\%$ & $98.70\%$ \\
				
				$b=0.9\omega/\tau_1$
				& $37.97\%$ & $97.00\%$ & $98.20\%$ \\
			\end{tabular*}
		\end{ruledtabular}
	\end{table}
	\section{Two-color dynamically assisted electric fields}\label{sec:DASE_fields}
	The second field we consider is the spatially homogeneous, time-dependent circularly polarized dynamically assisted electric field with frequency chirp. This is constructed by combining a strong-slowly varying field $\textbf{E}_{1s}(t)$ with a weak-rapidly varying field $\textbf{E}_{2w}(t)$, both sharing the circular polarization state, as follows:
	\begin{equation}\label{eq:DASE_field}
		\begin{split}
		\textbf{E}(t) & = \textbf{E}_{1s}(t) + \textbf{E}_{2w}(t)\\
		& = \frac{E_{1s0}}{\sqrt{2}}\,\text{exp}\left(-\frac{t^2}{2\tau^2}\right)\begin{pmatrix}
			\text{cos}(\omega_1 t + b_1 t^2)\\
			\text{sin}(\omega_1 t + b_1 t^2)\\
			0
		\end{pmatrix}\\
		&\hspace{1cm} + \frac{E_{2w0}}{\sqrt{2}}\,\text{exp}\left(-\frac{t^2}{2\tau^2}\right)\begin{pmatrix}
			\text{cos}(\omega_2 t + b_2 t^2)\\
			\text{sin}(\omega_2 t + b_2 t^2)\\
			0
		\end{pmatrix}.
		\end{split}
	\end{equation}
	Here, $E_{1s0}$ and $E_{2w0}$ denote the peak field strengths of the strong-slowly varying and weak-rapidly varying components, respectively, while $\tau$, $\omega_{1,2}$, and $b_{1,2}$ represent the pulse duration, field frequencies, and frequency chirps. We fix the field parameters as
	\begin{equation}\label{eq:DASE_field_parameters}
		\begin{split}
		&E_{1s0} = 0.2\sqrt{2}\text{E}_{cr}, \quad E_{2w0} = 0.25E_{1s0} = 0.05\sqrt{2}\text{E}_{cr}\\
		&\omega_1 = 0.1\,m,\quad \omega_2 = 8\omega_1 = 0.8\, m,\quad \tau = 40/m.
		\end{split}
	\end{equation}
	The purely time-dependent electric field in Eq.~\eqref{eq:DASE_field} can be viewed as the dipole approximation to a realistic space-time-dependent two-color standing-wave field generated by counterpropagating laser pulses. This approximation is expected to be valid when the characteristic spatial scales of the laser field, such as the focal width and carrier wavelength, are much larger than the pair-formation length scale \cite{aleksandrov2017momentum}. A fully space-time-dependent treatment is left for future work.
	\begin{figure}[tbh]
		\centering
		\includegraphics[width=\columnwidth]{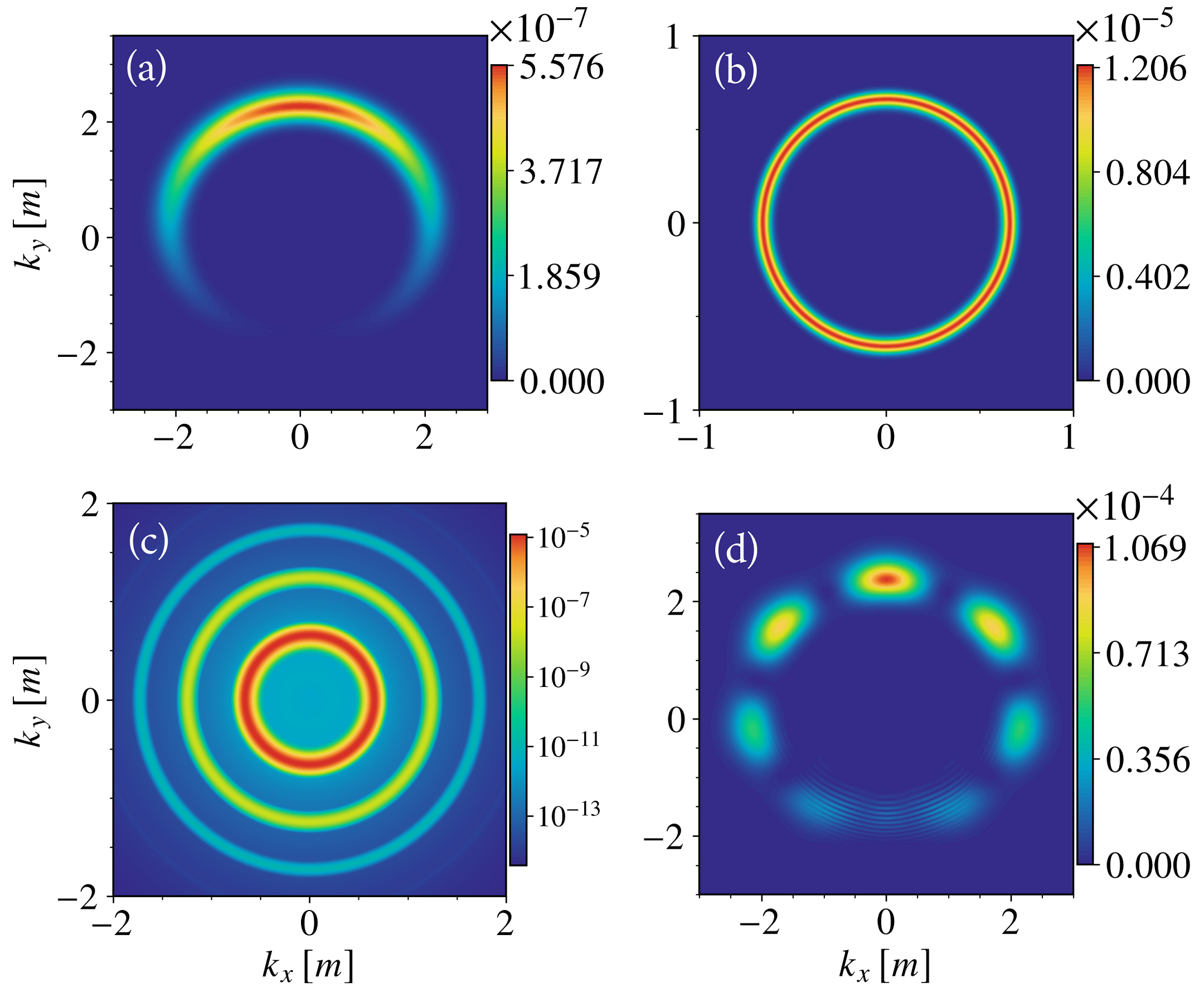}
		\caption{Helicity-summed momentum distributions of the created $e^+e^-$ pairs at $k_z = 0$ for (a) the one-color strong field $\textbf{E}_{1s}$, (b) the one-color weak field $\textbf{E}_{2w}$, (c) the one-color weak field $\textbf{E}_{2w}$ shown on a logarithmic scale, and (d) the combined two-color field $\textbf{E} = \textbf{E}_{1w} + \textbf{E}_{2w}$. All results correspond to the chirp-free case, $b_1=b_2=0$. The remaining field parameters are the same as those given in Eq. \eqref{eq:DASE_field_parameters}.}
		\label{fig:DASE_MD_(kx,ky)_chirp_free_fields}
	\end{figure}
	\begin{figure*}[tbh]
		\centering
		\includegraphics[width=\textwidth]{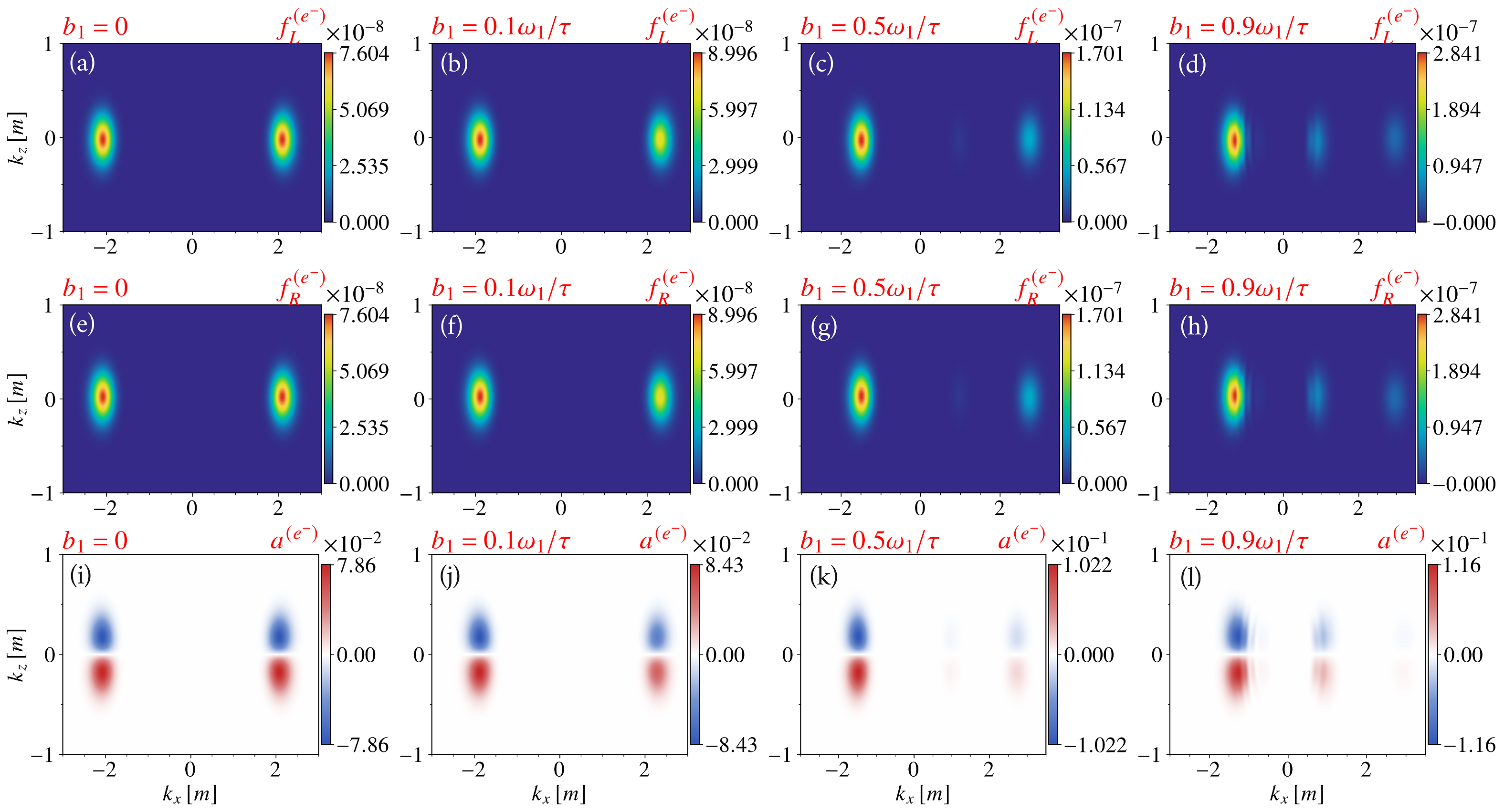}
		\caption{Helicity-resolved momentum distributions of the electrons produced in the one-color strong field $\textbf{E}_{1s}$, at $k_y=0$. Panels (a)--(d) show the left-helicity component $f_L^{(e^-)}$, panels (e)--(h) show the right-helicity component $f_R^{(e^-)}$, and panels (i)--(l) show the corresponding helicity asymmetry $a^{(e^-)}$. From left to right, the columns correspond to the chirp parameters $b_1[\omega_1/\tau]=0,0.1,0.5,$ and $0.9$, respectively. The remaining field parameter are the same as those given in Eq. \eqref{eq:DASE_field_parameters}.}
		\label{fig:DASE_MD_(kx,kz)_strong_field}
	\end{figure*}
	The above field configuration for arbitrary polarization was studied in detail in \cite{jangir2026dynamically}, with particular emphasis on helicity-summed momentum distributions and the interplay between polarization and chirping in dynamically assisted Schwinger pair production. Here, we revisit the circularly polarized configuration from a helicity-resolved perspective, focusing on the corresponding momentum distributions and helicity asymmetry of the produced particles.

	For the field parameters given in Eq. \eqref{eq:DASE_field_parameters}, the unchirped strong and weak fields have Keldysh parameters $\gamma_s=0.5$ and $\gamma_w=16$, respectively. In the following, we consider three chirping configurations: (i) strong-field chirping only, $b_1\neq0$, $b_2=0$; (ii) weak-field chirping only, $b_1=0$, $b_2\neq0$; and (iii) simultaneous chirping of both fields, with $b_1[\omega_1/\tau]=b_2[\omega_2/\tau]\neq0$. For each configuration, we present results for the individual strong and weak fields as well as for their two-color combination.

	Before discussing the helicity-resolved distributions for the field in Eq. \eqref{eq:DASE_field}, we first present the helicity-summed momentum distributions in the $(k_x,k_y)$ plane for the chirp-free case for completeness. The corresponding results are shown in Fig. \ref{fig:DASE_MD_(kx,ky)_chirp_free_fields}. The combined field exhibits the largest peak yield, followed by the strong and weak fields, demonstrating the effect of dynamical assistance. For the weak field [Fig. \ref{fig:DASE_MD_(kx,ky)_chirp_free_fields}(c)], the ring-like structures arise from multiphoton absorption processes \cite{blinne2016PhD, jangir2026dynamically}, with the visible rings corresponding to the absorption of three, four, and five photons. The multiple peaks in the combined-field distribution [Fig. \ref{fig:DASE_MD_(kx,ky)_chirp_free_fields}(d)] correspond to the characteristic shell structures reported in \cite{otto2015, olugh2020dynamically}.
	\subsection{Results for $\textbf{E}_{1s}(t)$}\label{subsec:DASE_strong_field_results}
	\begin{figure*}[tbh]
		\centering
		\includegraphics[width=\textwidth]{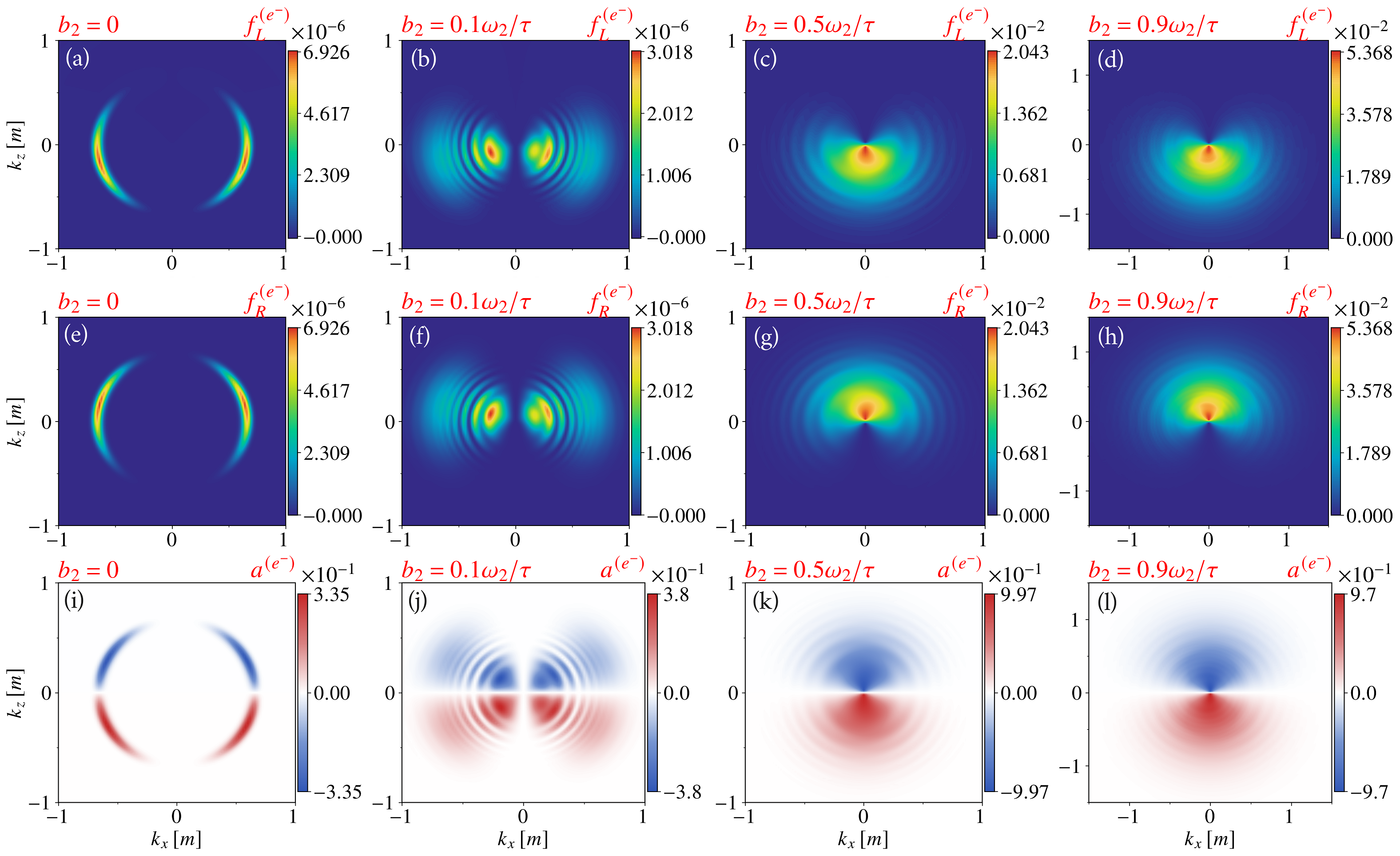}
		\caption{Helicity-resolved momentum distributions of the electrons produced in the one-color weak field $\textbf{E}_{2w}$, at $k_y=0$. Panels (a)--(d) show the left-helicity component $f_L^{(e^-)}$, panels (e)--(h) show the right-helicity component $f_R^{(e^-)}$, and panels (i)--(l) show the corresponding helicity asymmetry $a^{(e^-)}$. From left to right, the columns correspond to the chirp parameters $b_2[\omega_2/\tau]=0,0.1,0.5,$ and $0.9$, respectively. The remaining field parameter are the same as those given in Eq. \eqref{eq:DASE_field_parameters}.}
		\label{fig:DASE_MD_(kx,kz)_weak_field}
	\end{figure*}
	Figure \ref{fig:DASE_MD_(kx,kz)_strong_field} presents the helicity-resolved momentum distributions and helicity asymmetries of the produced electrons in the $(k_x$,$k_z)$ plane at $k_y=0$, in presence of $\textbf{E}_{1s}(t)$ for various chirp parameter $b_1$. For zero chirping ($b_1=0$), the momentum distributions exhibit two well-localized lobes centered at finite and opposite values of $k_x$, with an approximately symmetric profile. The two-lobe momentum structure is a consequence of the momentum acquired during pair production at different phases of the rotating electric field. As the chirp is introduced, the overall magnitude of the distribution changes, while the two-lobe structure is largely preserved. Nevertheless, a slight modification of the momentum localization and shape of the lobes becomes evident, particularly for the lobe at positive $k_x$. With increasing chirp strength, the peak yield grows progressively, and weak additional structures begin to develop in the region between the two dominant lobes, indicating an increasingly intricate momentum distribution.
	
	The left- and right-helicity components exhibit identical overall momentum support, but their distributions are separated with respect to the sign of $k_z$. The corresponding helicity asymmetry therefore displays a characteristic odd structure in $k_z$: the asymmetry is positive predominantly for negative $k_z$ and negative for positive $k_z$, indicating preferential production of left- and right-helicity electrons in the respective momentum regions. The magnitude of this helicity asymmetry increases gradually with the chirp, from about $8\%$ in the chirp-free case to approximately $11\%$ at the largest chirp considered. Thus, for the strong field alone, frequency chirping enhances both the overall pair-production yield and the helicity imbalance, while leaving the characteristic momentum-space structure largely intact.
	\subsection{Results for $\textbf{E}_{2w}(t)$}\label{subsec:DASE_weak_field_results}
	For the one-color weak field $\mathbf{E}_{2w}(t)$ alone, the helicity-resolved momentum distributions and the corresponding helicity
	asymmetry are shown in Fig.~\ref{fig:DASE_MD_(kx,kz)_weak_field}. In the chirp-free case ($b_2 = 0$), the distributions form two well-defined arc-like structures at opposite values of $k_x$, with the left- and right-helicity components exhibiting very similar momentum support. Even a small chirp substantially modifies this structure, producing pronounced oscillatory features around the two dominant momentum regions. With further increasing chirp, these oscillations evolve into a broad, centrally localized distribution, accompanied by a strong enhancement of the overall pair-production yield.
	
	The helicity asymmetry follows the same qualitative momentum-helicity correlation observed for the rotating field, with positive asymmetry predominantly in the negative-$k_z$ region and negative asymmetry for positive $k_z$. Its magnitude, however, increases strongly with the chirp. The maximum asymmetry grows from approximately $34\%$ in the chirp-free case to about $38\%$ for $b_2=0.1\omega_2/\tau$, and reaches nearly $100\%$ for $b_2=0.5\omega_2/\tau$ and $0.9\omega_2/\tau$. Thus, unlike the strong field, where the chirp produces only a modest change in the helicity imbalance, the weak field exhibits a pronounced chirp sensitivity, with the intermediate and large chirp regimes generating an almost complete helicity separation in momentum space.
	\begin{figure*}[tbh]
		\centering
		\includegraphics[width=\textwidth]{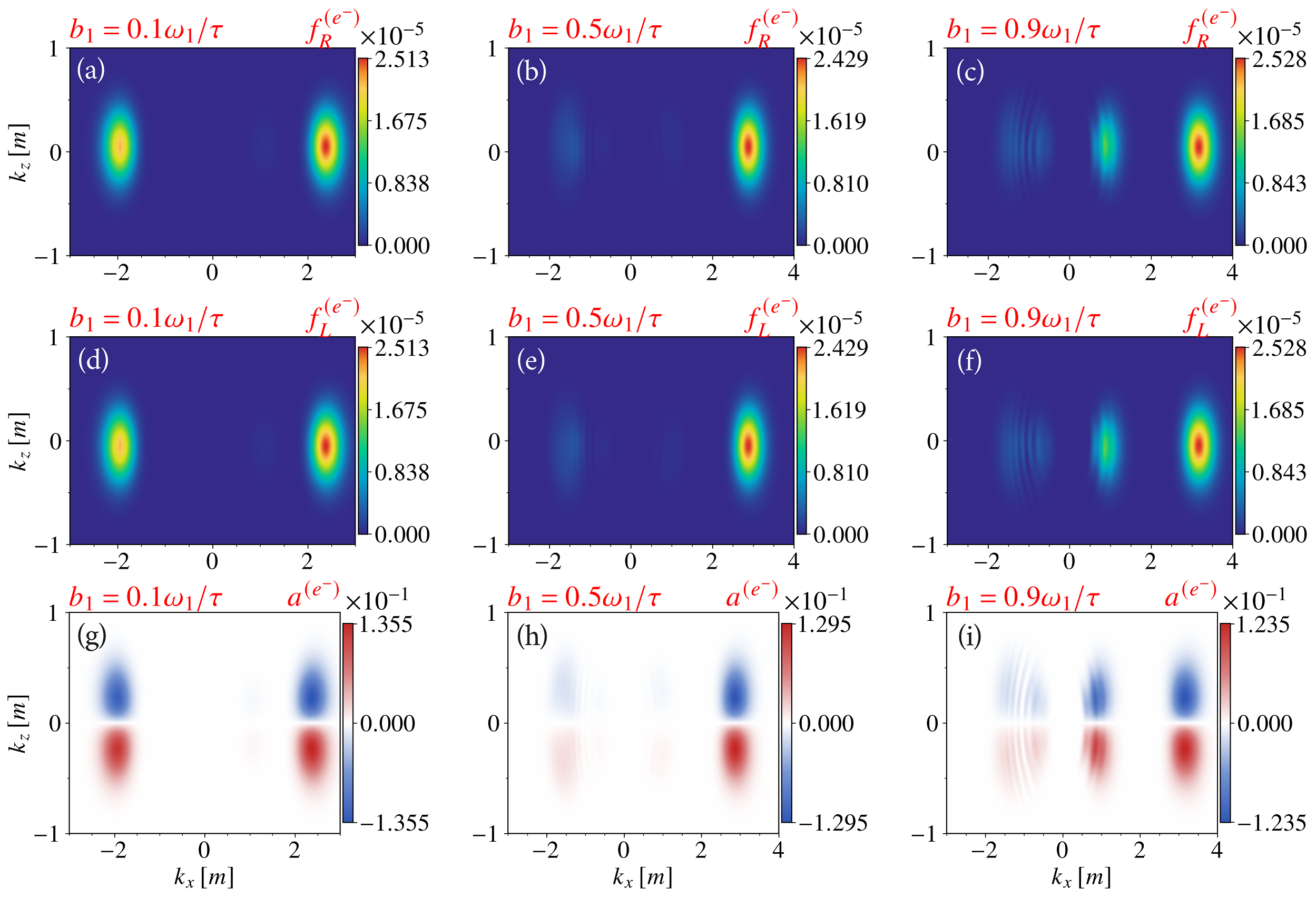}
		\caption{Helicity-resolved momentum distributions of the electrons produced in the two-color field $\textbf{E} = \textbf{E}_{1s} + \textbf{E}_{2w}$, at $k_y=0$. Panels (a)--(d) show the left-helicity component $f_L^{(e^-)}$, panels (e)--(h) show the right-helicity component $f_R^{(e^-)}$, and panels (i)--(l) show the corresponding helicity asymmetry $a^{(e^-)}$. From left to right, the columns correspond to the chirp parameters $b_1[\omega_1/\tau]=0.1,0.5,$ and $0.9$ applied to the strong-field component $\textbf{E}_{1s}$, while the weak-field component remains chirp-free, $b_2=0$. The remaining field parameter are the same as those given in Eq. \eqref{eq:DASE_field_parameters}.}
		\label{fig:DASE_MD_(kx,kz)_strong_chirps_combined_field}
	\end{figure*}
	\subsection{Results for $\textbf{E}(t) = \textbf{E}_{1s}(t) + \textbf{E}_{2w}(t)$}\label{subsec:DASE_combined_field_results}
	In this section, we present the results for dynamically assisted two-color field \eqref{eq:DASE_field} with various chirp scenarios. 
	\subsubsection{Chirp applied to $\textbf{E}_{1s}(t)$ only}\label{subsubsec:strong_field_chirp}
	For the two-color field $\mathbf{E}(t)=\mathbf{E}_{1s}(t)+\mathbf{E}_{2w}(t)$, with chirping applied only to the strong-field component ($b_1=0.1\omega_1/\tau$, $0.5\omega_1/\tau$, and $0.9\omega_1/\tau$) while $b_2=0$, the helicity-resolved momentum distributions and corresponding helicity asymmetry are presented in Fig. \ref{fig:DASE_MD_(kx,kz)_strong_chirps_combined_field}. The distributions preserve the characteristic two-lobe structure of the strong-field contribution, with the dominant peaks located at opposite values of $k_x$. As the chirp increases, weak oscillatory structures emerge between and around the dominant lobes, while the main momentum features remain well localized.
	
	Compared with the corresponding one-color strong-field case [Fig. \ref{fig:DASE_MD_(kx,kz)_strong_field}], the combined field exhibits a substantially larger momentum-space yield, demonstrating the enhancement associated with dynamical assistance. The main spectral features are only weakly affected by the strong-field chirp, whereas the subdominant structures become increasingly pronounced at larger chirp strengths.
	\begin{figure*}[tbh]
		\centering
		\includegraphics[width=\textwidth]{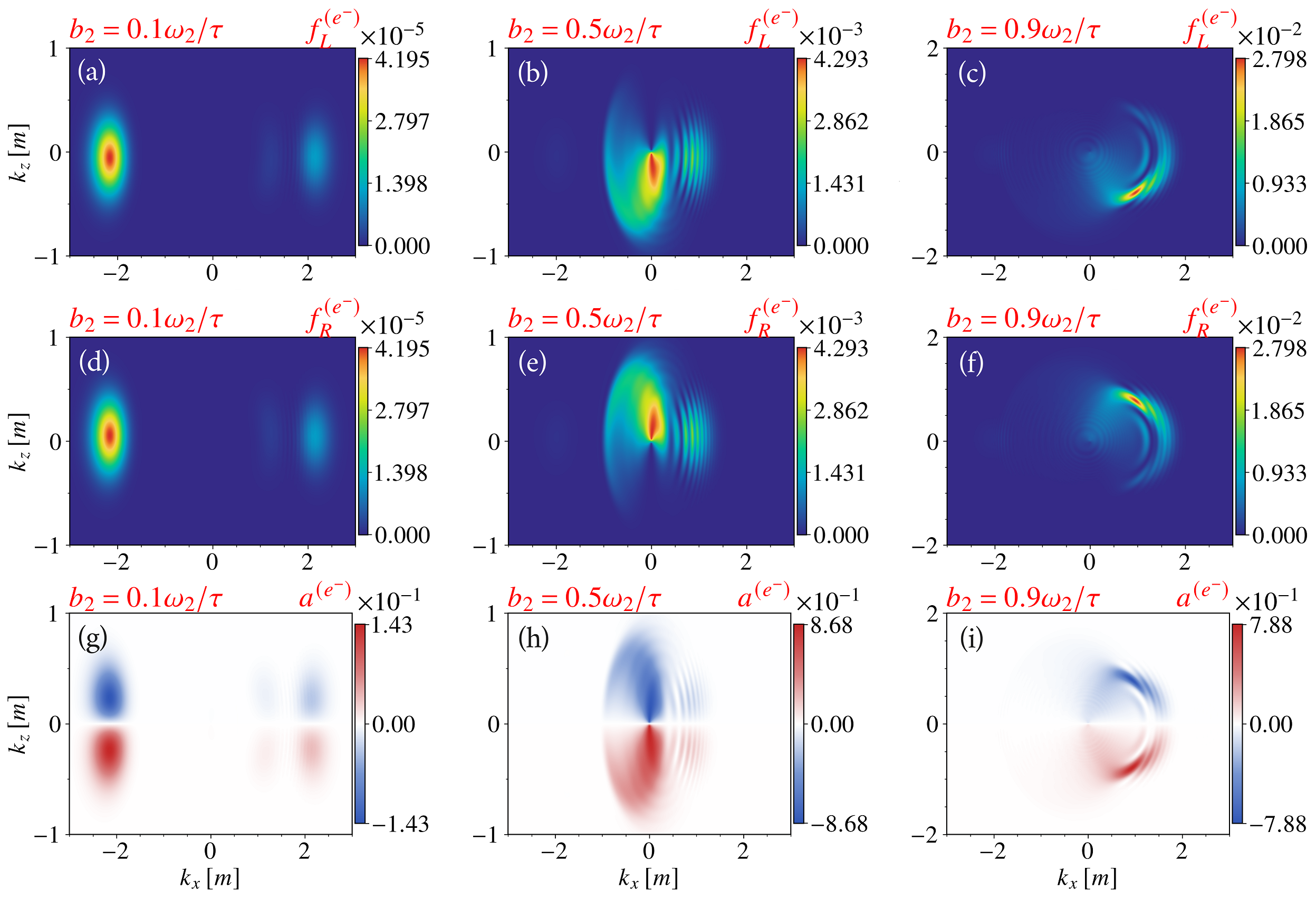}
		\caption{Helicity-resolved momentum distributions of the electrons produced in the two-color field $\textbf{E} = \textbf{E}_{1s} + \textbf{E}_{2w}$, at $k_y=0$. Panels (a)--(d) show the left-helicity component $f_L^{(e^-)}$, panels (e)--(h) show the right-helicity component $f_R^{(e^-)}$, and panels (i)--(l) show the corresponding helicity asymmetry $a^{(e^-)}$. From left to right, the columns correspond to the chirp parameters $b_2[\omega_2/\tau]=0.1,0.5,$ and $0.9$ applied to the weak-field component $\textbf{E}_{2w}$, while the strong-field component remains chirp-free, $b_1=0$. The remaining field parameter are the same as those given in Eq. \eqref{eq:DASE_field_parameters}.}
		\label{fig:DASE_MD_(kx,kz)_weak_chirps_combined_field}
	\end{figure*}
	
	The helicity asymmetry retains the characteristic odd dependence on $k_z$, with positive asymmetry predominantly for negative $k_z$ and negative asymmetry for positive $k_z$. Its maximum magnitude is approximately $13.6\%$, $13.0\%$, and $12.4\%$ for $b_1=0.1\omega_1/\tau$, $0.5\omega_1/\tau$, and $0.9\omega_1/\tau$, respectively. Thus, the helicity imbalance remains at the level of about $12$--$14\%$ and shows only a weak dependence on the strong-field chirp. In contrast to the one-color weak-field case [Fig. \ref{fig:DASE_MD_(kx,kz)_weak_field}], where chirping can generate a nearly complete helicity separation, the combined field considered here exhibits a much more moderate helicity asymmetry, indicating that the dominant strong-field contribution largely determines the helicity structure while the weak field substantially enhances the overall production yield and modifies the fine momentum-space structure.
	\subsubsection{Chirp applied to $\textbf{E}_{2w}(t)$ only}\label{subsubsec:weak_field_chirp}
	For the two-color field with chirping applied exclusively to the weak-field component, $b_2=0.1\omega_2/\tau$, $0.5\omega_2/\tau$, and $0.9\omega_2/\tau$, while the strong-field component remains chirp-free ($b_1 = 0$), the helicity-resolved momentum distributions and corresponding helicity
	asymmetries are shown in Fig.~\ref{fig:DASE_MD_(kx,kz)_weak_chirps_combined_field}. For the smallest weak-field chirp ($b_2 = 0.1\omega_2/\tau$), the momentum distributions retain the two-lobed structure inherited from the strong-field component. As $b_2$ is increased, the spectra undergo a pronounced reshaping: the dominant momentum structure becomes progressively broader, with the peak shifting toward positive $k_x$ and developing oscillatory, arc-like features. These deformations and the peak shift become most pronounced at the largest chirp.
	
	The peak yield increases strongly with the weak-field chirp. The enhancement is modest for $b_2=0.1\omega_2/\tau$, becomes several orders of magnitude stronger for $b_2=0.5\omega_2/\tau$, and increases further for $b_2=0.9\omega_2/\tau$. Thus, chirping the weak high-frequency field has a much stronger influence on the overall pair-production yield than chirping the strong field alone, particularly in the intermediate- and large-chirp regimes.
	\begin{figure*}[tbh]
		\centering
		\includegraphics[width=\textwidth]{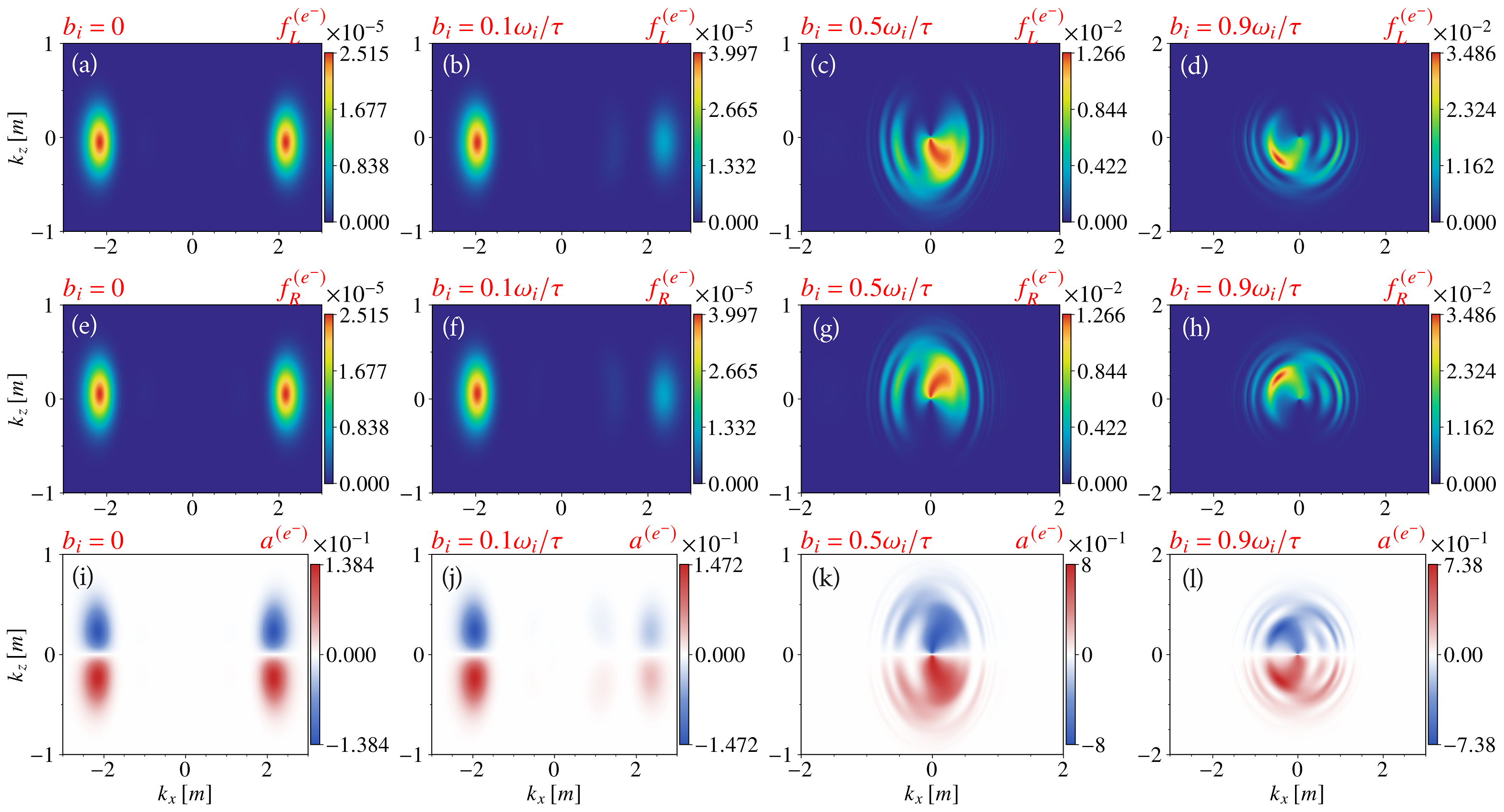}
		\caption{Helicity-resolved momentum distributions of the electrons produced in the two-color field $\textbf{E} = \textbf{E}_{1s} + \textbf{E}_{2w}$, at $k_y=0$. Panels (a)--(d) show the left-helicity component $f_L^{(e^-)}$, panels (e)--(h) show the right-helicity component $f_R^{(e^-)}$, and panels (i)--(l) show the corresponding helicity asymmetry $a^{(e^-)}$. From left to right, the columns correspond to the chirp parameters $b_i[\omega_i/\tau]=0,0.1,0.5,$ and $0.9$, with $i=1,2$, where the chirps are applied simultaneously to the strong and weak field components while maintaining the same normalized chirp strength for both components. The remaining field parameter are the same as those given in Eq. \eqref{eq:DASE_field_parameters}.}
		\label{fig:DASE_MD_(kx,kz)_strong_weak_chirps_combined_field}
	\end{figure*}
	
	The helicity asymmetry also responds strongly to the weak-field chirp. Its maximum magnitude is approximately $14\%$, $87\%$, and $79\%$ for $b_2=0.1\omega_2/\tau$, $0.5\omega_2/\tau$, and $0.9\omega_2/\tau$, respectively. Thus, the helicity imbalance increases sharply with chirp strength, peaking at nearly $87\%$ at the intermediate chirp before decreasing slightly at the largest chirp. The asymmetry remains predominantly positive for negative $k_z$ and negative for positive $k_z$, although the increasingly structured momentum spectra at larger $b_2$ produce corresponding oscillatory features in the asymmetry.

	Compared with the case in which the chirp is applied to the strong field only [Fig. \ref{fig:DASE_MD_(kx,kz)_strong_chirps_combined_field}], the weak-field chirp produces a substantially stronger helicity response. In the former case, the asymmetry remains at approximately $12$--$14\%$ over the considered chirp range, whereas chirping the weak field raises it to tens of percent and ultimately to nearly $90\%$. This demonstrates that the high-frequency weak component plays a particularly important role in controlling the helicity structure of the two-color pair-production process.
	\subsubsection{Chirp applied to both $\textbf{E}_{1s}(t)$ and $\textbf{E}_{2w}(t)$}\label{subsubsec:both_field_chirp}
	For simultaneous chirping of both field components, with $b_i[\omega_i/\tau]=0.1$, $0.5$, and $0.9$ for $i=1,2$, the helicity-resolved momentum distributions and the corresponding helicity asymmetry are shown in Fig.~\ref{fig:DASE_MD_(kx,kz)_strong_weak_chirps_combined_field}. In the chirp-free case ($b_i=0$), the momentum distributions exhibit two well-separated and dominant structures, characteristic of the combined field and similar to those observed for the strong-field component alone [Fig. \ref{fig:DASE_MD_(kx,kz)_strong_field}]. For a small simultaneous chirp ($b_i=0.1\omega_i/\tau$), the locations of the dominant structures remain essentially unchanged. However, the peak in the positive-$k_x$ region becomes slightly more diffuse, accompanied by a reduction in its peak value. As the chirp strength increases to $b_i=0.5\omega_i/\tau$ and $0.9\omega_i/\tau$, the distributions undergo a pronounced restructuring, with broader arc-like and oscillatory features developing around the central momentum region. Along with these spectral modifications, the overall pair-production yield increases substantially with increasing chirp strength.
	\begin{table*}[t]
		\centering
		\caption{Maximum helicity asymmetry, expressed as a percentage, for the dynamically assisted electric field $\mathbf{E}(t)$ defined in Eq. \eqref{eq:DASE_field}, along with its strong-field component $\mathbf{E}_{1s}(t)$ and weak-field component $\mathbf{E}_{2w}(t)$. Here, $a_i=b_i[\omega_i/\tau]$ (with $i = 1,2$) denotes the dimensionless chirp strength. For the combined field, the three configurations correspond to simultaneous chirping of both components ($b_1\neq0$, $b_2\neq0$), chirping of the strong-field component only ($b_1\neq0$, $b_2=0$), and chirping of the weak-field component only ($b_2\neq0$, $b_1=0$).}
		\label{tab:max_helicity_asymmetry}
		\renewcommand{\arraystretch}{1.35}
		\begin{ruledtabular}
			\begin{tabular*}{\textwidth}{@{\extracolsep{\fill}}llcccc}
				Field & Chirped component(s)
				& $a_i=0$ & $a_i=0.1$ & $a_i=0.5$ & $a_i=0.9$ \\
				\hline
				$\mathbf{E}_{1s}(t)$ & Strong field ($b_1$)
				& $7.86\%$ & $8.43\%$ & $10.22\%$ & $11.60\%$ \\
				
				$\mathbf{E}_{2w}(t)$ & Weak field ($b_2$)
				& $33.50\%$ & $38.00\%$ & $99.70\%$ & $97.00\%$ \\
				
				\hline
				$\mathbf{E}(t)$ & Strong and weak fields
				& $13.84\%$ & $14.72\%$ & $80.00\%$ & $73.80\%$ \\
				
				$\mathbf{E}(t)$ & Strong field only
				& $-$ & $13.55\%$ & $12.95\%$ & $12.35\%$ \\
				
				$\mathbf{E}(t)$ & Weak field only
				& $-$ & $14.30\%$ & $86.80\%$ & $78.80\%$ \\
			\end{tabular*}
		\end{ruledtabular}
	\end{table*}
	The helicity asymmetry follows the same odd-in-$k_z$ pattern, with positive asymmetry predominantly in the negative-$k_z$ region and negative asymmetry in the positive-$k_z$ region. Its magnitude is about $13.8\%$ for the chirp-free case and remains nearly unchanged at $14.7\%$ for the small simultaneous chirp. A pronounced enhancement occurs at the intermediate chirp, where the asymmetry reaches about $80\%$. For the largest chirp, the maximum asymmetry decreases slightly to approximately $74\%$, while retaining the same overall momentum-helicity correlation. The oscillatory structures in the momentum distributions are also reflected in the spatially structured helicity-asymmetry pattern at the larger chirps.
	
	A comparison of the three chirping configurations for the combined field $\mathbf{E}(t)$, summarized in Table \ref{tab:max_helicity_asymmetry}, shows that chirping the weak-field component alone produces the strongest helicity response, with a maximum asymmetry of about $87\%$ at intermediate chirp strength. In contrast, chirping the strong-field component alone results in only a weak response, with the maximum asymmetry remaining in the range of $12\%$--$14\%$. Simultaneous chirping of both components also enhances the asymmetry, but its enhancement is slightly weaker than that obtained by chirping the weak-field component alone at the corresponding chirp strengths. This behavior is qualitatively consistent with the weak-field case, where the helicity asymmetry increases strongly at intermediate and large chirp strengths, reaching $99.7\%$ and $97\%$ for intermediate and large chirp strengths, respectively. These results indicate that the weak-field chirp plays the dominant role in enhancing the helicity asymmetry of the combined field, whereas chirping the strong-field component alone has only a minor effect.
	\section{Conclusion and Discussion}\label{sec:conclusion}
	In this work, we have investigated helicity signatures in nonperturbative electron--positron pair production driven by spatially homogeneous, time-dependent circularly polarized electric fields with frequency chirping. Within the quantum-kinetic equation formalism for fermions, we have calculated the helicity-resolved momentum distributions and characterized the resulting helicity dependence through the helicity asymmetry. We have considered different field configurations, including temporally asymmetric and dynamically assisted fields, and systematically examined how frequency chirping modifies the helicity-resolved momentum distributions and influences the magnitude of the helicity asymmetry.
	
	The helicity-resolved momentum distributions are presented in the $(k_x,k_z)$ plane at $k_y=0$, with the electric field confined to the $(x,y)$ plane. Across all the field configurations considered, we observe a clear helicity-dependent structure in the produced electron distributions. The helicity asymmetry is predominantly odd in the out-of-plane momentum $k_z$, indicating a preferential separation of the two helicity components along this direction. In particular, left-helicity electrons are predominantly enhanced at negative $k_z$, whereas right-helicity electrons are preferentially enhanced at positive $k_z$. Although the two helicity components occupy largely overlapping regions of the $(k_x,k_z)$ momentum space, their distributions exhibit a clear separation with respect to the sign of $k_z$. Consequently, the helicity asymmetry approximately changes sign under $k_z\rightarrow-k_z$, giving rise to its characteristic odd dependence on the out-of-plane momentum.
	
	In the case of temporally asymmetric electric fields, the influence of frequency chirping becomes increasingly pronounced as the temporal asymmetry of the pulse is varied. For weak chirping, the dominant momentum peaks remain localized near the same out-of-plane momentum positions, while additional oscillatory structures begin to appear, particularly for the elongated pulses. At intermediate chirp strengths, the momentum distributions become broader and more structured, and the helicity asymmetry can increase to nearly $100\%$ for the elongated pulses. At the largest chirp strength, the momentum distributions undergo substantial reshaping and broadening, accompanied by large helicity asymmetries of about $97\%$--$98\%$ for the symmetric and elongated pulses, respectively. In contrast, the compressed-pulse case behaves differently, with its helicity asymmetry decreasing at large chirp strengths and the dominant helicity--$k_z$ correlation being reversed. Thus, the helicity response in temporally asymmetric fields is governed by a nontrivial interplay between pulse asymmetry and frequency chirping, rather than by either parameter independently.
	
	We next investigated dynamically assisted pair production using a two-color circularly polarized field consisting of a strong, low-frequency component and a weak, high-frequency component. The helicity-summed distributions demonstrate the characteristic enhancement associated with dynamical assistance: the combined field produces a substantially larger yield than either component considered separately, while the weak field exhibits ring-like momentum structures associated with multiphoton processes. The helicity-resolved analysis reveals that the two components do not contribute equally to the helicity response. For the one-color strong field, chirping mainly enhances the yield and modifies the fine momentum structure without producing a large change in the helicity imbalance, whereas, in the case of the one-color weak field, it contributes much less to the pair yield, and its frequency chirp has a disproportionately strong effect on the helicity structure of the produced particles, with the helicity asymmetry reaching nearly $100\%$ at the intermediate and largest chirp strengths considered. For the two-color dynamically assisted field, we examined three distinct chirping configurations: chirping of the strong field alone, chirping of the weak field alone, and simultaneous chirping of both components with the same normalized chirp strength. The results show that chirping the weak field alone produces the strongest helicity response, reaching nearly $87\%$ at intermediate chirp strength, whereas chirping the strong field alone results in only a weak response, with the maximum asymmetry remaining in the range of $12\%$--$14\%$. Simultaneous chirping combines these effects and leads to both substantial spectral restructuring and a large helicity imbalance, although the resulting helicity asymmetry is weaker than that obtained by chirping the weak field alone.
	
	In summary, these results establish helicity-resolved momentum distributions as a sensitive observable of the temporal structure and frequency content of rotating electric fields and demonstrate that frequency chirping of the high-frequency assisting field provides an efficient means of controlling the helicity composition of particles produced from the quantum vacuum. Our results provide further insight into the role of helicity in nonperturbative vacuum pair production and may help guide future investigations of helicity-dependent signatures in strong-field QED experiments.
	\bibliography{ref} 
\end{document}